# Ferroelectric brightening of spin-forbidden dark excitons in a $WSe_2$/hybrid-perovskite heterostructure

Xinyun Wang[1,2†], Magdalena Grzeszczyk[3†], Maxim Trushin[3,4,5†], Ivan Verzhbitskiy[6,7], Dmitrii Litvinov[3], Yi Wei Ho[2,3], Yuan Chen[1], Zhenyue Wu[1], Mykola Telychko[1], Chuanqi Zhang[1], Andres Granados del Aguila[3], Kuan Eng Johnson Goh[2,6,7,8], Xinwei Li[2], Goki Eda[1,2,4], Shaffique Adam[2,4], Maciej Koperski[3]*, and Kian Ping Loh[1]*

[1]Department of Chemistry, National University of Singapore, Singapore, Singapore.

[2]Department of Physics, National University of Singapore, Singapore, Singapore.

[3]Institute for Functional Intelligent Materials, National University of Singapore, Singapore, Singapore.

[4]Centre for Advanced 2D Materials and Graphene Research Centre, National University of Singapore, Singapore, Singapore.

[5]Department of Materials Science Engineering, National University of Singapore, Singapore, Singapore.

[6]Quantum Innovation Center, Agency for Science Technology and Research, Singapore, Singapore.

[7]Institute of Materials Research and Engineering, Agency for Science Technology and Research, Singapore, Singapore.

[8]Division of Physics and Applied Physics, School of Physical and Mathematical Sciences, Nanyang Technological University, Singapore, Singapore.

[†]Equal contribution to this work.

## Abstract

Long-lived dark excitons in monolayer $WSe_2$ present promising candidates for carrying spin and valley information, but their optical access and spin manipulation have conventionally required the use of strong external magnetic fields. Here, using a ferroelectric hybrid perovskite heterostructure, we leverage the ferroelectric proximity effect to break the $WSe_2$'s in-plane rotational symmetry and brighten the spin-forbidden dark excitons under zero magnetic field conditions. Furthermore, we show that the twist angle between the $WSe_2$ and perovskite crystals controls the ferroelectric coupling strength and valley-contrasting polarization. Our proposed mechanism, supported by a four-band tight-binding model, suggests that the ferroelectric proximity effect induces an asymmetric intersublattice interaction, generating an effective in-plane spin-orbit coupling (SOC) field that rotates spin/valley polarization and brightens dark excitons. Our work establishes ferroelectric proximity coupling as an electrically reconfigurable, magnetic-field-free strategy for spin exciton control in two-dimensional semiconductors.

## Main Text

In monolayer $WSe_2$, the band-edge exciton is theoretically predicted and experimentally confirmed to be spin-forbidden and optically dark. Due to short-range exchange interactions, this dark exciton manifold splits into two distinct states: a strictly dark exciton ($X^D$) and a grey exciton ($X^G$)[1-4]. These states form as coherent superpositions of valley-polarized configurations, where $X^D$ is antisymmetric and completely dipole-forbidden, remaining optically inactive under out-of-plane detection, while $X^G$ is symmetric and possesses a small out-of-plane dipole moment, granting it limited but nonzero oscillator strength[5,6]. Because of their suppressed radiative decay, these long-lived dark and grey excitons serve as protected reservoirs for valley and spin information. Their optical access and spin state control, enabled via magnetic fields or tailored optical pulses, make monolayer $WSe_2$ a highly promising material for valleytronics and quantum information processing.

Dark excitons have been activated most directly by in-plane magnetic fields that mix bright and dark states, revealing strong emissions and long valley lifetimes[7-9]. Magnetic-free routes, such as near-field plasmonic coupling[10,11] and tip-enhanced nano-cavity control[12], intensify dark-exciton emission by coupling to the out-of-plane dipole moment and increasing the radiative rate through strong Purcell enhancement. These methods generally require sophisticated structural design and rely on enhancing weak radiative pathways ($X^G$) rather than changing the excitonic spin orientation. In this context, electric-field-assisted SOC routes are appealing as it is more feasible for on-chip device integration. However, conventional Rashba SOC induced by an out-of-plane electric field is intrinsically weak, owing to the small Rashba coefficient around band extrema. As a result, appreciable spin mixing requires either very large local electric fields or alternative symmetry-breaking mechanisms[13-15].

Here, we demonstrate the magnetic-free brightening of dark exciton ($X^D$) in monolayer $WSe_2$, enabled by ferroelectric proximity and its intrinsic Kane-Mele (KM) SOC, realized by interfacing $WSe_2$ with the in-plane ferroelectric hybrid perovskite $(TMA)_3Sb_2Cl_9$ (TSC). Both the ferroelectric brightening and the polarization-phase difference of the dark/grey exciton doublet are tunable with the interlayer twist angle. Our theoretical analysis indicates that the directional in-plane ferroelectric field couples to the $WSe_2$ electronic wavefunctions and generates a sublattice-asymmetric spin-orbit interaction, producing an effective in-plane SOC field that rotates the exciton spin configuration and activates the otherwise spin-forbidden transition. This mechanism is supported by the observed symmetry reduction, the twist-angle-dependent magneto-photoluminescence (PL) response, and the polarization phase shift of the dark/grey exciton doublet. Together, these results establish ferroelectric proximity as a symmetry-tunable, magnetic-free route to static spin and exciton control in two-dimensional quantum materials.

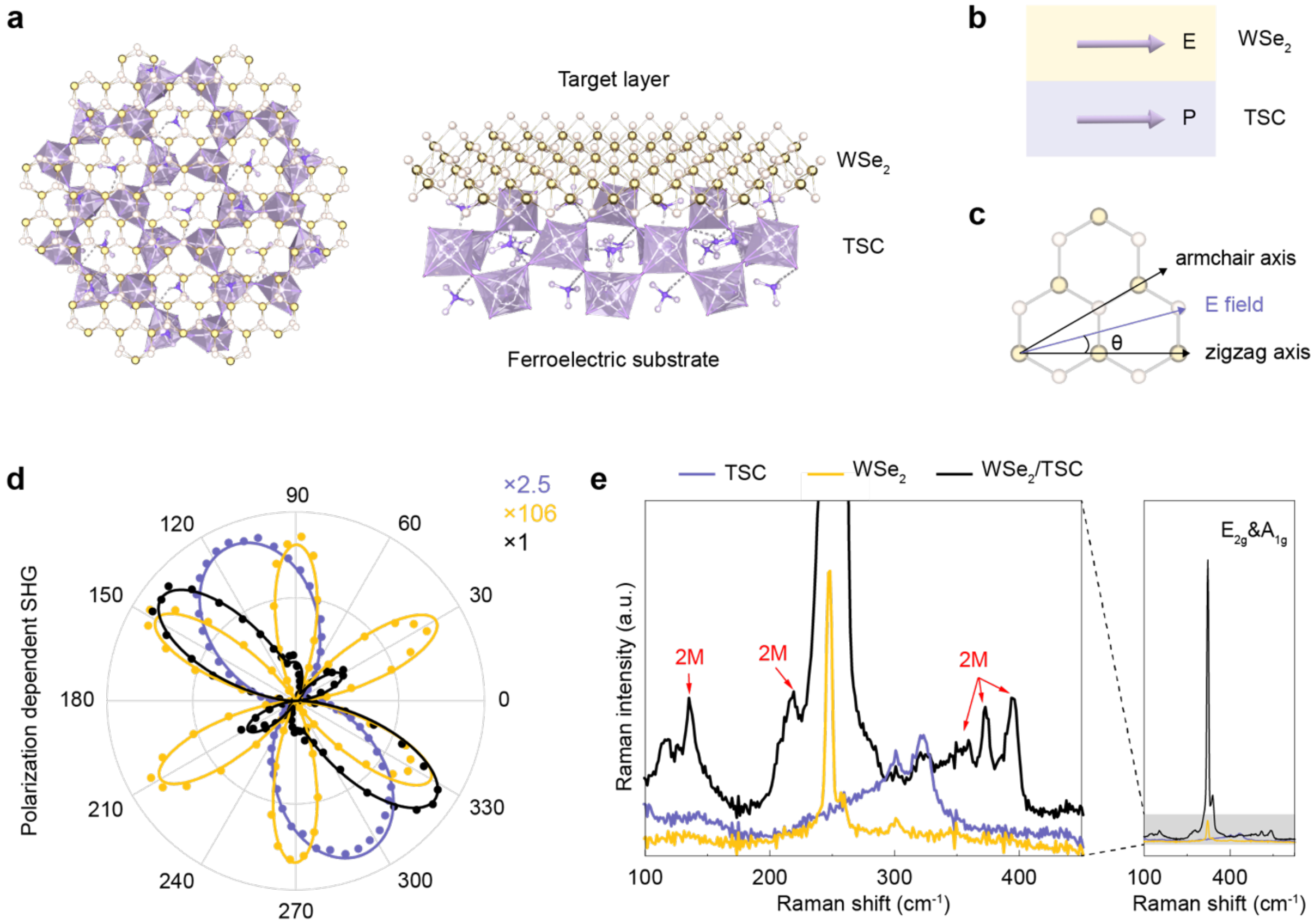

**Fig. 1 | Symmetry breaking in $WSe_2$ via ferroelectric proximity coupling. a**, Schematic illustration of the heterostructure: monolayer $WSe_2$ on a 2D ferroelectric perovskite (TSC). **b**, The ferroelectric polarization $P$ imprints a microscopic in-plane electric field $E$ across the interface. **c**, The twist angle $\theta$ is defined between the $WSe_2$ zigzag axis and the TSC ferroelectric polarization. **d**, Polar plot of co-polarized SHG response from the individual crystals and the heterostructure at a twist angle $\theta$ of 6.7°. The monolayer $WSe_2$ shows a six-lobe polar pattern, whereas the $WSe_2$/TSC heterostructure exhibits a strongly asymmetric pattern that cannot be achieved by linear superposition of the individual crystal tensors. Fits require $C_1$ symmetry for $WSe_2$, indicating the three-fold rotational and horizontal mirror-symmetry breaking (see Supplementary Section 1 for fitting details). **e**, Raman scattering spectra from the same regions. The heterostructure displays additional features (marked as 2M) consistent with relaxation of Raman selection rules under reduced symmetry.

Ferroelectric proximity coupling is implemented in a vertical heterostructure (Fig. 1a; fabrication details in Methods). The in-plane ferroelectric polarization in TSC produces bound surface charges and an associated electrostatic potential that extends into the neighbour $WSe_2$ layer, effectively "imprinting" a directional in-plane electric field onto $WSe_2$ (Fig. 1b). The motivation for choosing TSC as the ferroelectric substrate is twofold: (1) Its dangling-bond-free surface forms atomically clean van der Waals interface with $WSe_2$, enabling high sample quality and efficient field transfer without chemical hybridization; (2) Its large bandgap makes it as an optical transparent substrate and avoid interfacial charge transfer during light excitation. Detailed characterization of TSC is available in in our previous work[16] and Supplementary Fig. S1. Importantly, the interlayer twist angle offers an additional control knob, allowing the proximity field to align preferentially with either the zigzag or armchair directions of $WSe_2$ (Fig. 1c).

Linearly co-polarized second-harmonic generation (SHG) measurement is first used to determine the stacking angle of the heterostructure by examining the crystallographic orientation of individual layers. In accordance with previous reports, monolayer $WSe_2$ exhibits six symmetric lobes (Fig. 1d) under normal incidence, reflecting its three-fold rotational symmetry ($D_{3h}$ point group)[17,18]. The TSC crystal shows a two-lobe polarization pattern, corresponding to its $C_s$ symmetry. After fitting the SHG polar map from individual layers, a interfacial twist-angle of 6.7° is determined for this heterostructure.

Notably, in this heterostructure region, the obtained SHG polar map cannot be fitted using any linear superposition of the respective susceptibility tensor of $D_{3h}$ and $C_s$ symmetry point group. We further tested each subgroup under the $D_{3h}$ symmetry and found that only the lowest $C_1$ point group can fit the experimental data. Due to Neumann's principle, the crystal symmetry is encoded in the second-order nonlinear susceptibility tensor $\chi^{(2)}$, determining the efficiency and occurrence of the SHG process. Therefore, the reduced $\chi^{(2)}$ tensor indicates the broken three-fold rotational and horizontal mirror reflection symmetry of $WSe_2$ in the heterostructure. A detailed SHG analysis is described in Supplementary Section 1.

Furthermore, Fig. 1d reveals several second-order and combination Raman modes, typically activated only in strained $WSe_2$[19-22], which are clearly present in the heterostructure. The emergence of these forbidden or silent modes (2M) signifies a relaxation of Raman scattering selection rules and evidences symmetry breaking. Note that these 2M modes also arise when $WSe_2$ is interfaced with other ferroelectric substrates, indicating that the reduced symmetry originates from $WSe_2$, rather than from the ferroelectric substrate (see Supplementary Fig. S2).

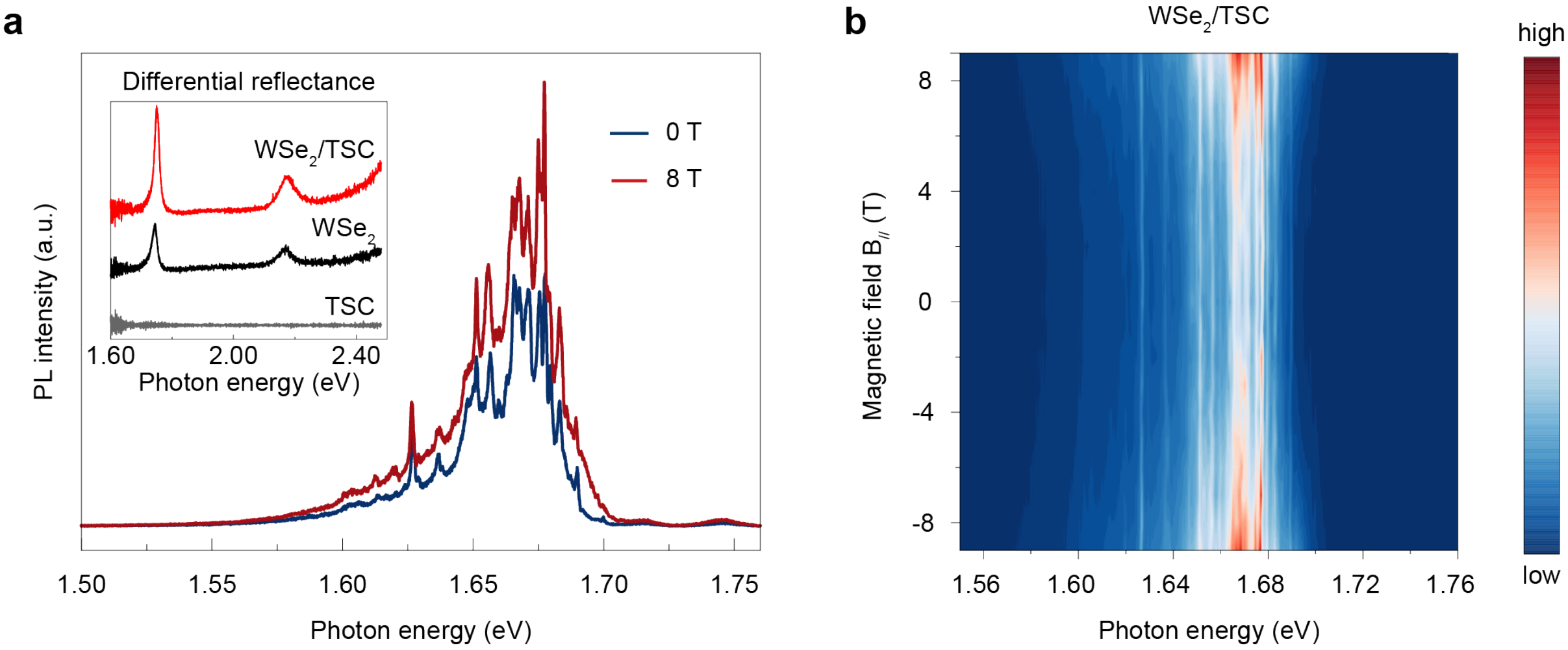

**Fig. 2 | Observation of spin-forbidden dark excitons in $WSe_2$/TSC heterostructures at zero magnetic field. a**, PL spectra of a $WSe_2$/TSC heterostructure at 0 T and 8 T. Inset is the differential-reflectance spectra of TSC perovskite, $WSe_2$ monolayer, and $WSe_2$/TSC heterostructure. **b**, False-color map of magneto-PL from a $WSe_2$/TSC heterostructure as a function of $B_{//}$.

Fig. 2a shows the PL spectra of the stacked $WSe_2$/TSC heterostructure, where a new group of excitonic peaks emerges. No excitonic resonances were observed from bare TSC (inset of Fig. 2a), consistent with its large optical bandgap (>3.5 eV). Moreover, the differential reflectance spectrum of the heterostructure closely resembles that of pristine monolayer $WSe_2$ and is dominated by A and B excitons[23,24], indicating that TSC acts as an optically transparent substrate and the new PL peaks are associated with $WSe_2$.

Importantly, the PL emission from these new states intensifies with an increasing in-plane magnetic field ($B_{//}$) (Fig. 2b). As demonstrated in previous studies[3,7-9], the application of $B_{//}$ can rotate the spin polarization through interband wavefunction mixing and thereby brighten spin-forbidden dark excitonic states. Consequently, the magnetic-field-induced brightening provides a definitive signature of the spin-forbidden nature of these newly observed excitonic states.

The dark nature is further corroborated by the g-factor extracted from out-of-plane magnetic-field ($B_{\perp}$) dependent PL measurements in a valley-resolved configuration[25-29] (see Supplementary Fig. S3). While the $WS_2$/TSC heterostructure exhibits similar emergent excitonic peaks, no such counterpart excitonic states are observed in the $MoSe_2$/TSC heterostructure (Supplementary Fig. S4). This material selectivity can be explained by the fact that the lowest exciton is exclusively dark in tungsten-based TMDs[30-33]. Finally, comprehensive gate-, power-, and magnetic-field-dependent measurements (Supplementary Figs. S5–S7) exclude alternative origins such as trions, many-body complexes, and defect-bound excitons, thereby confirming that the emergent peaks originate from neutral spin-forbidden excitons.

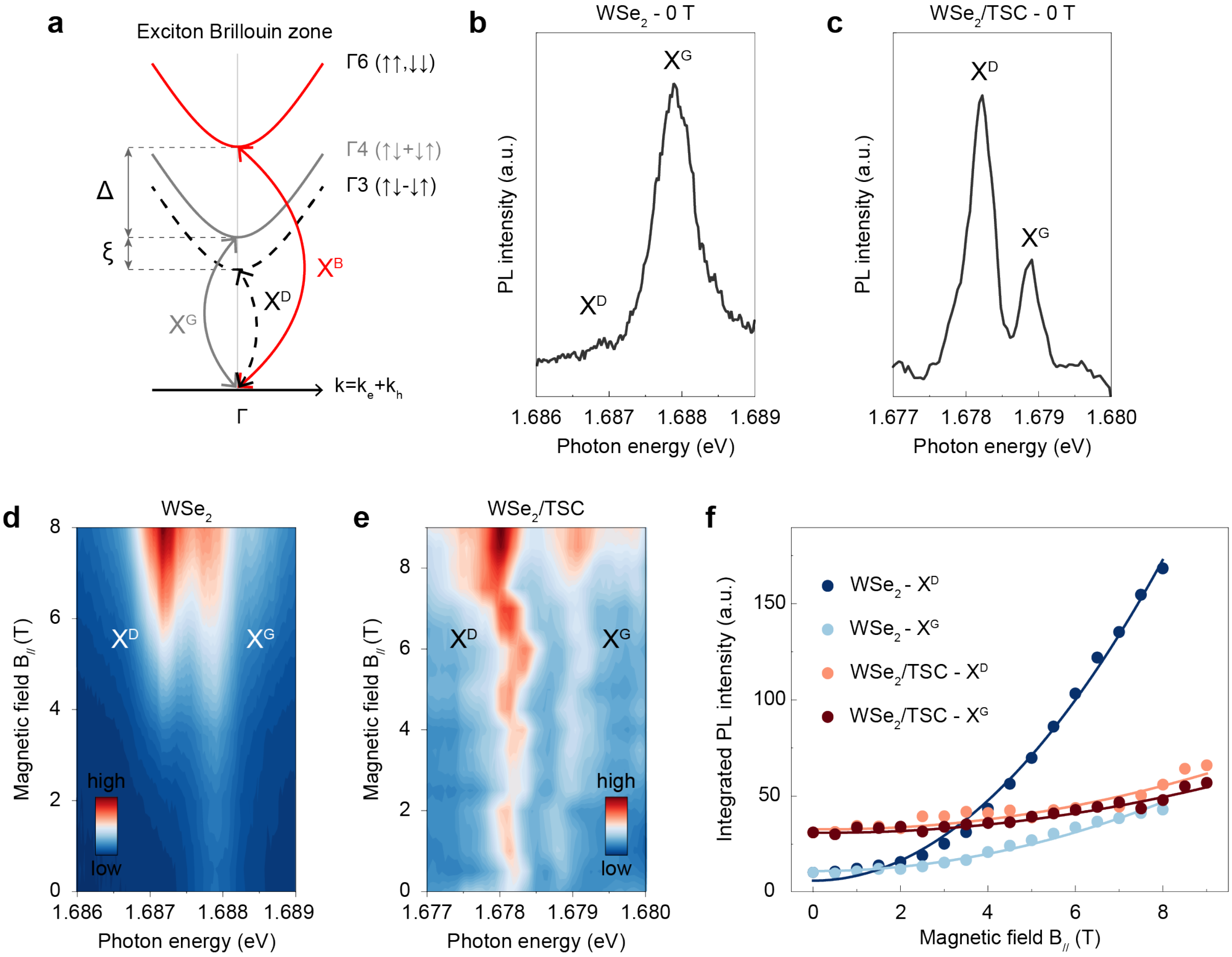

**Fig. 3 | Magnetophotoluminescence signatures of dark-grey exciton doublet. a**, Exciton fine structure of a pristine $WSe_2$ monolayer (point group $D_{3h}$). The bands are labeled with the irreducible representations of the $D_{3h}$ point group. Δ is the spin splitting from on-site SOC and $\xi$ is the fine splitting from the short-range exchange interaction. Antiparallel spins at the band edges render the lowest neutral exciton $X^D$ optically dark, while the higher energy state $X^G$ acquires a finite out-of-plane dipole moment and is optically accessible. Irreducible-representation labels follow $D_{3h}$. **b**,**c**, High-resolution PL spectra of a monolayer

and a 6.7° stacked heterostructure at zero magnetic field. **d**,**e**, Corresponding magneto-PL map (in-plane magnetic field $B_{//}$). **f**, Integrated intensities of $X^D$/$X^G$ doublets as a function of $B_{//}$ field. The parabolic functions ($I = I_0 + \alpha B_{\parallel}^2$) are employed for the fitting. The observation of a smaller curvature $\alpha$ for the heterostructure is consistent with the pre-brightening effect induced by ferroelectric proximity.

Next we turn to the fine structure of the dark manifold and examine whether the newly observed peaks are associated with the grey exciton state. Fig. 3a illustrates the exciton fine structure of a pristine $WSe_2$ monolayer. In monolayer $WSe_2$, this doublet can be resolved under an external in-plane magnetic field at higher energy resolution, exhibiting a splitting energy of ~663 μeV[3,5] (Fig. 3b). At zero magnetic field, $X^D$ is absent, while $X^G$ may be observed in high-quality samples, characterized by in-plane emission direction and attributed to the large numerical aperture of the objective lens. In this 6.7° stacked heterostructure, the doublet fine structures can also be resolved for each dark state. Fig. 3c highlights the dominant exciton doublet at 1.678-1.679 eV. Importantly, both $X^D$ and $X^G$ exhibit intensive emission at zero magnetic field.

Similar to the magneto-response of the pristine monolayer, the exciton doublet in heterostructure shows further enhancement under applied $B_{//}$ fields, confirming their spin-forbidden dark nature (Fig. 3d,e). The spectral wandering of the doublet can be attributed to fluctuations of the ferroelectric electric field, giving rise to a jittering effect (Supplementary Fig. S8). Furthermore, in both systems the emission intensities of the $X^G$ and $X^D$ excitons increase parabolically with the in-plane magnetic field, following $I = I_0 + \alpha B_{\parallel}^2$ (ref[3,4];Fig. 3f). The smaller proportionality coefficient ($\alpha$) in the heterostructure reflects reduced magnetic-field-induced spin mixing compared to the pristine monolayer, consistent with a pre-brightening effect driven by ferroelectric proximity.

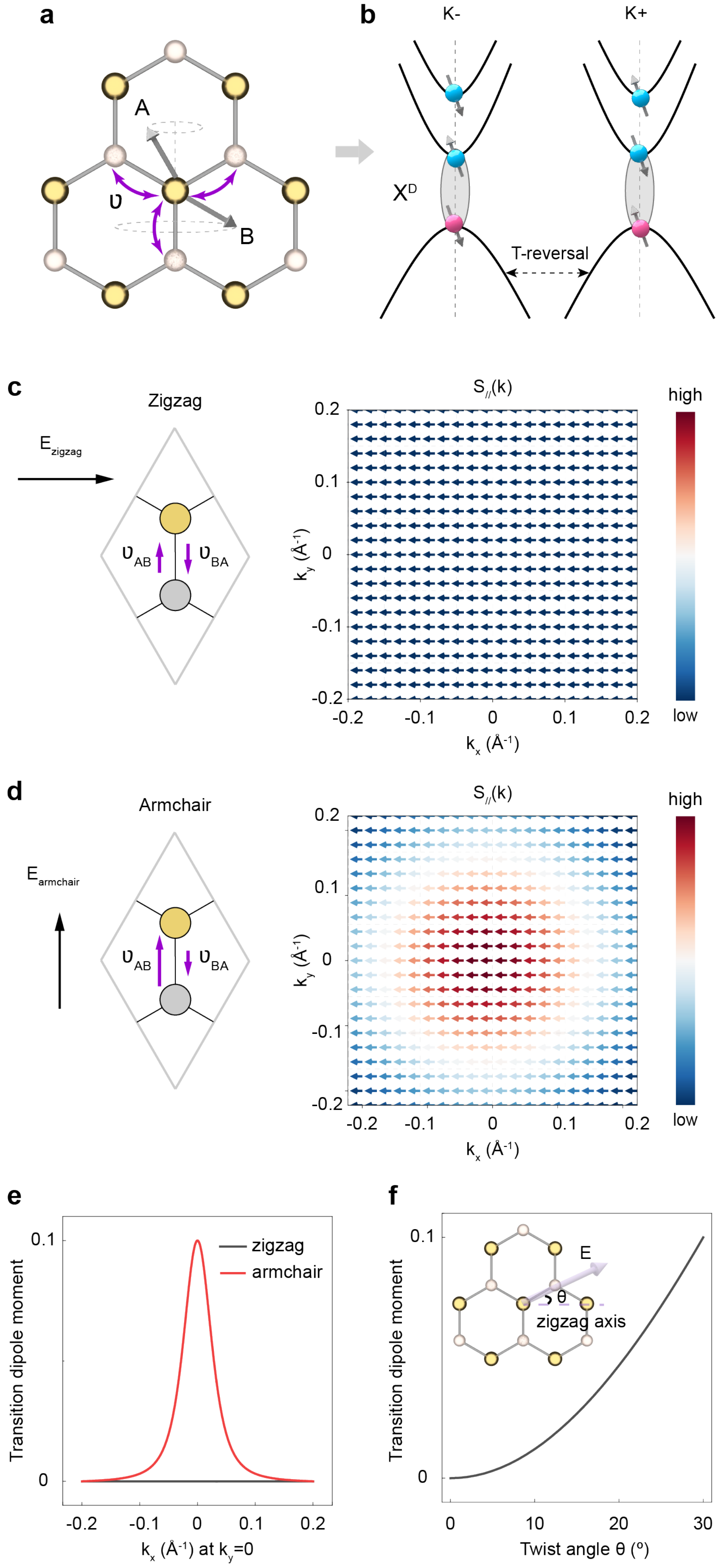

a
A
υ
B
b
K-
K+
$X^D$
T-reversal
c
Zigzag
$E_{zigzag}$
$υ_{AB}$
$υ_{BA}$
$S_{//}(k)$
high
low
$k_y$ (Å$^{-1}$)
$k_x$ (Å$^{-1}$)
d
Armchair
$E_{armchair}$
e
Transition dipole moment
zigzag
armchair
$k_x$ (Å$^{-1}$) at $k_y$=0
f
E
θ
zigzag axis
Twist angle θ (°)

**Figure 4 | Theoretical investigation of spin canting induced by asymmetric intersublattice interaction.** **a**, Schematic diagram of the spin-orbit-coupled honeycomb lattice with resolved sublattices. The spin reorientation arises from KM-type SOC modified by the asymmetrical hopping between the two sublattice sites. **b**, Schematic illustrates the spin canting at band edges. **c**,**d**, Calculated quiver maps of in-plane spin components at the lowest conduction band of $WSe_2$ in zigzag- and armchair-aligned heterostructure. Arrows indicate spin orientation and color encodes magnitude. **e**, Calculated $k$-dependent transition dipole moment. **f**, Calculated twist-angle-dependent transition dipole moment at $k = 0$. The inset illustrates the twist angle ($\theta$) between the zigzag axis of $WSe_2$ and the in-plane ferroelectric field.

We now propose a mechanism for dark-exciton brightening based on the ferroelectric proximity coupling. In honeycomb lattices, KM-type SOC couples real spin to the sublattice pseudospin and establishes an Ising-like spin texture at the band edges[34-36]. As a result, the spin orientation and the associated optical selection rules are highly sensitive to perturbations that differentiate the two sublattices (occupied by W and Se atoms in $WSe_2$) and modify inter-sublattice interaction[24,37,38] In the $WSe_2$/TSC heterostructure, the in-plane ferroelectric polarization of TSC generates a directional interfacial electric field ($E$) onto the $WSe_2$ lattice. This field lowers $WSe_2$ symmetry and modifies the intersublattice electrostatic potential, leading to a SOC-enabled spin rotation (Fig. 4a). In a tight-binding picture, this imposes an asymmetry between nearest-neighbor A→B and B→A hopping amplitudes ($v_{AB} \neq v_{BA}$). At the band edge, this hopping asymmetry could manifest as an effective in-plane SOC field, which reorients the band-edge spins and relaxes the spin-selection rule (Fig. 4b).

Next we develop a four-band Hamiltonian to capture the ferroelectric proximity effect, in which the $E$ field enters as a direction-dependent perturbation term and determines the effective intersublattice hopping asymmetry. In particular, we focus on two representative alignment cases: the zigzag-aligned and armchair-aligned configurations, where the $E$ field is perpendicular and parallel to the $WSe_2$ lattice, respectively, as illustrated in Fig. 4c,d. All other alignment configurations can be interpreted as linear combinations of these two cases. Notably, for zigzag-aligned case, the net effective field along the lattice bond are identical due to the intrinsic $C_3$ symmetry and the nearest-neighbor intersublattice hoppings yield symmetric amplitudes ($v_{AB} = v_{BA}$), thus the effective in-plane SOC field is negligible (Fig. 4c). In contrast, for armchair-aligned case, the intersublattice electrostatic potential is altered and the hopping strength become asymmetric ($v_{AB} \neq v_{BA}$), giving rise to an in-plane effective SOC field and produces a pronounced local maximum at $k = 0$ (Fig. 4d). The corresponding in-plane spin textures of the topmost valance band is shown in Supplementary Fig. S9.

Fig. 4e displays the calculated $k$-dependent transition dipole moment, which vanishes for zigzag alignment but shows a pronounced increase for armchair alignment. This result is consistent with our experimental observations in multiple devices tested: the magneto-response is absent only in the zigzag-aligned heterostructure but appears in all other alignment configurations. The twist-angle dependent transition dipole moment (Fig. 4f) reveals a more specific correlation to the ferroelectric field projection. The monotonic dependence reveals that the spin rotation is proportional to the $E$ field projection along armchair axis. In addition, we found that the transition dipole moment is independent of the magnitude of the on-site spin-flip term and is only determined by the hopping asymmetry, which is not considered by previous studies[6,13]. In fact, a finite on-site spin-flip term produces a uniform in-plane spin texture for both the lowest conduction band and highest valence band, with equally uniform magnitude but opposite sign. Consequently, the spins remain antiparallel leaving the interband transition spin-forbidden. Detailed analysis is presented in Supplementary Section 2.

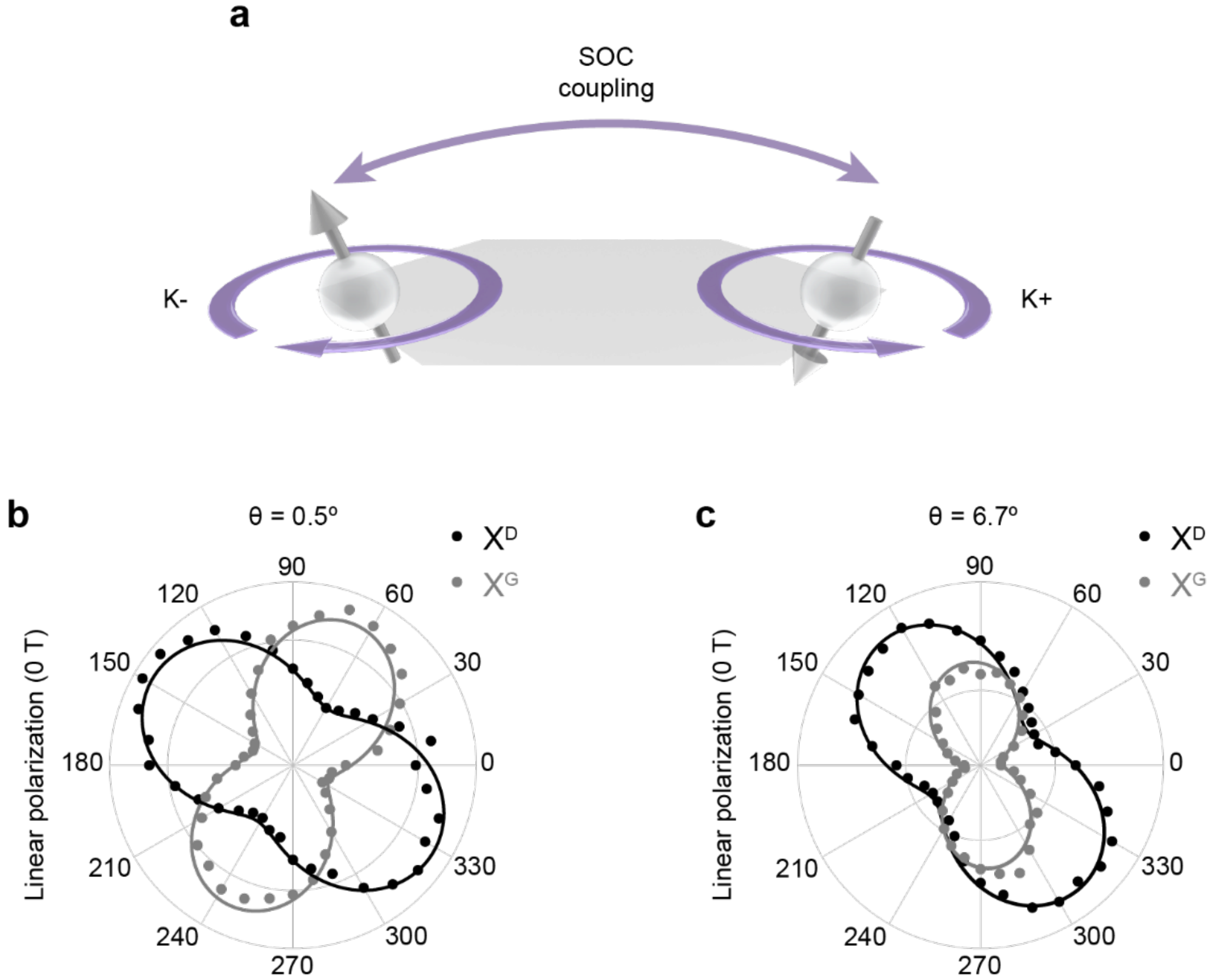

**Fig. 5 | Twist-angle dependent phase difference. a**, Schematic of SOC-induced valley mixing. Oppositely polarized spins couple to the time-reversal-related $K^{\pm}$ valleys, in which the valley polarization is encoded in the photon helicity (purple arrows). **b**,**c**, Linear polarization-resolved emission of the dark-grey exciton doublet for heterostructures with a twist angle of 0.5° and 6.7° at 0 T.

The ferroelectric proximity mechanism is further supported by a reduction in the dark/grey exciton polarization phase difference. Unlike PL intensity, which can vary with extrinsic factors such as carrier doping and defects, the polarization orientation is an intrinsic property, arising from the coherent superposition of the two valley exciton states (Fig. 5a), and it therefore directly tracks the degree of intervalley coherence (valley mixing)[5,6,39]. In pristine monolayers, the doublet polarization exhibits a 90° phase difference, reflecting two orthogonal valley-coherent superpositions[3,39,40]. In the heterostructures (Fig. 5b,c), this phase difference is reduced from 90° to 82.1° and 30.5° for twist angles of 0.5° and 6.7°, respectively, indicating a rotation of the valley-coherent basis. This observation follows naturally from the coupling between valley pseudospin and sublattice pseudospin: near $K^{-}$ and $K^{+}$, the in-plane sublattice pseudospin texture carries opposite phase winding and encodes opposite valley chirality.

Consequently, any perturbation that distinguishes A from B introduces an effective intervalley-coupling term in the valley-coherent subspace, mixing the two valley components and rotating the polarization basis of the dark/grey doublet.

The observed phase reduction thus provides a robust fingerprint of our tight-binding theoretical model above, where the effective SOC field as well as the related spin and valley mixing degree is twist-angle dependent, vanishing at zigzag-aligned configuration and maximized at armchair-aligned configuration. Although a systematic twist-angle dependence is limited by the demand for high-quality samples to resolve darkmanifolds, the two representative heterostructures studied here are sufficient to distinguish the expected angular trend, following our calculated phase diagram based on the group theory (Supplementary Section 2.3).

## Conclusion

In this study, we demonstrate the modulation of exciton spin and valley polarization via an interfacial ferroelectric proximity effect. This ferroelectric proximity is achieved in a van der Waals heterostructure combining a ferroelectric perovskite substrate with a monolayer $WSe_2$. At zero magnetic field, we observe the brightening of spin-forbidden dark-grey excitons in $WSe_2$, as confirmed by their distinct magneto-optical response. We explain this effect using the KM-type SOC model, where the intersublattice hopping strength is regulated by the orientation of the ferroelectric field. This mechanism is additionally supported by the magnetic response and polarization properties of the dark/grey exciton doublet. Overall, our findings establish ferroelectric proximity coupling as an effective, magnetic-field-free strategy for manipulating spin and pseudospin degrees of freedom, an approach readily extendable to other low-dimensional quantum systems.

# Methods

**Sample fabrication.**

Because of the instability of perovskite in air, an all-dry transfer method is used to fabricate the devices in a glove box with oxygen and moisture levels below 0.01 parts per million. First, TMD monolayers are obtained by micromechanical cleavage of bulk TMD crystal (HQ graphene) onto a 285 nm $SiO_2$/Si substrate. hBN crystals were exfoliated onto a poly(dimethylsiloxane) (PDMS) stamp and then used to pick up TMD monolayers. The hBN thickness is typically 20–40 nm. Second, TSC crystals are grown in between the sapphire or Si substrates inside a vacuum chamber at room temperature for 24 hrs. (Detailed structural and ferroelectric characterization data can be found in our previous work[16] and Supplementary Fig. S1.) Third, the pre-prepared hBN-TMD stack is brought into contact with TSC crystal carefully. During the transfer process, both mechanical force and heat are minimized to achieve a high-quality interface. In addition, the quality of the cleaved surface of TSC crystal is double-checked by PFM before transfer (details of PFM and atomic AFM studies are provided in SI). For gate devices, few-layer graphene flakes are exfoliated on PDMS stamp for dry transfer and used as top and contact gate electrodes. After fabrication, heterostructures are covered by hexagonal boron nitride (hBN) crystals to prevent environmental degradation during the sample loading process.

**Optical measurements.**

Photoluminescence (PL) measurements were conducted using two cryostats: the attoDRY2100 and the attoDRY800, operating at base temperatures of 1.6 K and 4.2 K, respectively. The attoDRY2100 cryostat was equipped with a superconducting coil capable of generating magnetic fields up to 9 T. Sample cooling in the attoDRY2100 was achieved using helium exchange gas, while the attoDRY800 used thermal contact with a cold finger for cooling. The sample was mounted on a chip carrier positioned on x/y piezo-positioners for precise alignment. An in-situ objective with a numerical aperture of 0.82 was fixed to a $z$ piezo-positioner, enabling accurate focalization of laser light onto the device surface. The optical signal was collected and dispersed using a 0.75 m spectrometer equipped with gratings of 150, 600, and 1800 g/mm. Detection was performed using a liquid nitrogen-cooled charge-coupled device (CCD) camera. Magnetic states were probed with the magnetic field applied both in-plane and out-of-plane relative to the c-axis of $WSe_2$. A fiber-based probe was used for these measurements. Polarization-resolved experiments (detection) were conducted with the help of polarizers and quarter-wave plates. The excitation energy was tuned close to the bright exciton emission resonance using a supercontinuum source, with an excitation power of 10 μW focused on the sample. Gated structures were characterized using a Keithley 2400 source meter.

Differential reflectance (DR) measurements were performed using a laser confocal system with an ST-500 Janis cryostat at 10 K. A tungsten-halogen lamp was used as the white light source. The light was focused through a lens and directed into an objective lens (Olympus SLMPLN 100×, NA 0.6), producing a beam with a size of approximately 1 μm. The reflected signal was collected and analyzed using the Andor Solis spectrometer equipped with iDus Si CCD.

The second harmonic generation (SHG) measurement was carried out at room temperature in a home-built multiphoton microscope, pumping with an ultrafast laser (Toptica FemtoFiber Pro NIR, 1550 nm, 90 fs, 80 MHz) through an objective lens (Olympus LCPLN100XIR). The signal was collected in the back-scattering geometry with the same lens. For polarization-resolved SHG, co-polarized excitation and detection beam are used in a normal incidence configuration. A rotating achromatic half-waveplate (Thorlabs SAHWP05M-700) was used to control both the pump laser and signal linear polarization angles, before analyzed with a fixed polarizer and detected by a spectrometer (Andor Shamrock 500i and iKon L BV).

Raman measurements were conducted at room temperature using a WITec alpha300R confocal Raman microscope system equipped with 532 nm laser excitation. A 50× objective lens was used to focus the laser spot on the sample surface to a diameter of around 1 μm. A spectral grating with 1800 lines/mm was used for collecting Raman spectra.

# Acknowledgements

X.Wang acknowledges valuable discussions with prof. Jiangbin Gong and prof. Yuanping Feng from NUS physics department. X.Wang sincerely thanks Dr. Qi Zhang, Dr. Ziying Wang, Dr. Junyong Wang, Dr. Zehua Hu, Dr. Guangwei Hu and Dr. Hao Wang for the experimental suggestions and helps at the early stage of this project. K. P. Loh acknowledges support from the Ministry of Education (MOE), Singapore, under AcRF Tier 3 (MOET32024-0001). M. Koperski acknowledges the support from the Ministry of Education (MOE), Singapore, through the Research Centre of Excellence program (grant EDUN C-33-18-279-V12, I-FIM), the Ministry of Education (MOE), Singapore, under its Academic Research Fund Tier 2 (MOE-T2EP50122-0012) and the Air Force Office of Scientific Research and the Office of Naval Research Global under award number FA8655-21-1-7026. M. Trushin. is supported by the Ministry of Education (MOE), Singapore, under its Research Centre of Excellence award to the Institute for Functional Intelligent Materials (I-FIM, project No. EDUNC-33-18-279-V12). X. Li. acknowledge the support from the Ministry of Education (MOE), Singapore, through Award MOE-T2EP50124-0005. I. Verzhbitskiy and K. E. J. Goh acknowledge the support from the Agency for Science, Technology and Research (A*STAR) (#21709, C230917006, C230917007), as well as Singapore National Research Foundation (CRP21-2018-0001).

## Author contributions

X. Wang and K. P. Loh conceived the idea and initiated this project. X. Wang carried out the sample preparations. M. Grzeszczyk, I. Verzhbitskiy, D. Litvinov, Y. Chen, A. G. D. Aguila, and X. Wang performed PL measurements. Y. W. Ho and X. Wang performed SHG and Raman measurements. M. Telychko and C. Zhang performed ncAFM characterization. X. Wang performed PFM characterization. M. Trushin and S. Adam devised the tight-binding model. M. Trushin and X. Wang performed theoretical calculation. X. Wang and X. Li performed group theory analysis. K. P. Loh and M. Koperski supervised the project. X. Wang analyzed the experimental data and wrote the manuscript. All authors discussed the results and commented on the manuscript.

## Competing interests

The authors declare no competing interests.